\documentclass[sts]{imsart}

\RequirePackage[OT1]{fontenc}
\usepackage{amsthm,amsmath,natbib}
\RequirePackage[colorlinks,citecolor=blue,urlcolor=blue]{hyperref}

\usepackage{amssymb,epsf, here, amsfonts,latexsym,graphicx,pictexwd,natbib}

\newcommand{\FF}{{\mathbb F}}

\newcommand{\RR}{{\mathbb R}}

\usepackage{graphicx}

\startlocaldefs
\numberwithin{equation}{section}
\theoremstyle{plain}

\endlocaldefs

\begin{document}

\begin{frontmatter}
\title{A Conversation with Jon Wellner} 
\runtitle{Interview with Jon A. Wellner}

\begin{aug}
\author{\fnms{Moulinath} \snm{Banerjee}\thanksref{t1}\ead[label=e1]{moulib@umich.edu}} \and
\author{\fnms{Richard} J. \snm{Samworth}\thanksref{t2}\ead[label=e2]{r.samworth@statslab.cam.ac.uk}}

\thankstext{t1}{Moulinath Banerjee is Professor of Statistics and Biostatistics at the University of Michigan, Ann Arbor, MI \printead{e1}.}
\thankstext{t2}{Richard J. Samworth is Professor of Statistical Science and Director of the Statistical Laboratory at the University of Cambridge, UK \printead{e2}.}
\runauthor{M. Banerjee and R. J. Samworth}

\affiliation{University of Michigan at Ann Arbor and University of Cambridge}


\end{aug}
\bigskip

\begin{abstract}
Jon August Wellner was born in Portland, Oregon, in August 1945.  
He received his Bachelor's degree from the University of Idaho in 1968 and his PhD 
degree from the University of Washington in 1975.  
From 1975 until 1983 he was an Assistant Professor and Associate
Professor at the University of Rochester.  
In 1983 he returned to the University of Washington, and has remained
at the UW as a faculty member since that time.  Over the course of a long and distinguished career, Jon has made seminal contributions to a 
variety of areas including empirical processes, semiparametric theory, and shape-constrained inference, and has co-authored a number 
of extremely influential books. He has been honored as the Le Cam lecturer by both the IMS (2015) and the French Statistical Society (2017).
He is a Fellow of the IMS, the ASA, and the AAAS, and an elected member of the International Statistical Institute. He has served as co-Editor of 
Annals of Statistics (2001--2003) and Editor of Statistical Science (2010--2013), and President of IMS (2016--2017). In 2010 he was made a Knight of the Order of the Netherlands Lion. In his free time, Jon enjoys mountain climbing and backcountry skiing in the Cascades and British Columbia.
\end{abstract}

\maketitle
\end{frontmatter}
\newpage
\begin{description}
\item[Mouli:] 
Jon, I want to say that it's a privilege to interview you. 
I would like to thank you for being a fantastic advisor and for your guidance, 
encouragement and inspiration throughout the years. \\
\item[Richard:]	
To echo what Mouli says, it's a great pleasure for me too! 
\end{description}
	
\section{Childhood}

\begin{description}
\item [Mouli:] 
Can you tell us a bit about your family and background, and in particular, your 
father Charles Wellner, to whose work you have a link on your webpage? 
In particular, how influential was he in fostering your love of nature and the outdoors? \\

\item[Jon:]  I was born in Portland, Oregon, and grew up in Missoula (Montana), Spokane (Washington),
and Ogden (Utah). My father was a research forester; he started his career doing white-pine silviculture, and 
ended as a research administrator, organizing research centers for forestry in conjunction with 
Universities in the intermountain west:  Utah State (Logan), Montana State (Bozeman),  
and the U. of Idaho (Moscow).
He had a huge influence on my interests in outdoor activities.  
He was a skier during the 1920's and 1930's before organized ski areas were developed.
\bigskip


\item[Richard:]  Did your interests in quantitative fields start emerging during your school 
years? Any specific memories of school-life that you would like to share with us? 

\item[Jon:] During high-school in Ogden (Utah), I did not focus on studies, 
but I did spend quite a bit of time skiing at the local ski area (Snow Basin).  
When I began undergraduate work at the U of Idaho, I was initially interested in pursuing a 
career in forestry, but started enjoying mathematics during my undergraduate work.
So I switched majors after three years and ended up with  
Bachelors degrees in Math and Physics.

\item[Mouli:] Forestry's loss was Statistics' gain! 
\bigskip

\begin{figure}[ht!]
\centering
\includegraphics[width=110mm]{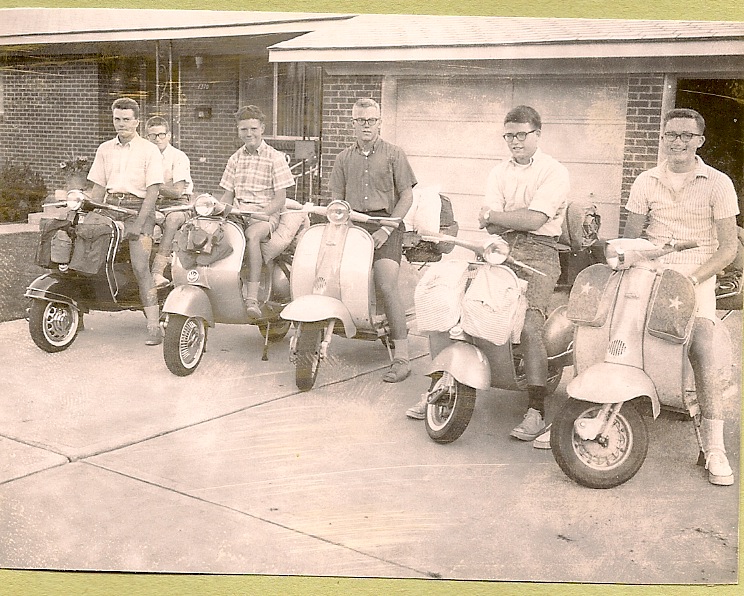}
\caption{The intrepid and inimitable Vespa Gang. Which one is Jon?}
\label{fig:figure1}
\end{figure}

\item[Mouli:] We remember a picture of you during your 65th birthday celebration as part of a `scooter-gang'. 
Can you tell us a bit about that? 

\item[Jon:]  Several of my friends during high school had scooters, and I also bought a small Vespa to 
get to work at an ice cream factory in Ogden.  The scooter gang (Robert Johnstone, Steve Keller, 
and others)  organized several longer summer trips during those years:  
once to Cedar Breaks National Monument 
in Utah, and a longer trip to Yellowstone and Grand Teton National parks.   
I remember bucking a terrific head-wind in Wyoming during the return trip from Yellowstone.

\item[Richard:] What a great photo!
\end{description}

\section{Undergraduate years}
\begin{description}
\item[Richard:] You majored in Mathematics and Physics at University of Idaho. 
How was your academic experience there as an undergraduate? Did you consider other schools? 

\item[Jon:] I started at the U of I  in September 1963 after graduating from high school in Ogden in June 1963.  
My parents had both attended the U of I during the 1930's, so it was a natural place to go in terms of 
family history.  The U of I also had a scholarship incentive program that made it very affordable:  
if you earned a  certain GPA (about 3.7 or above), then out of state students would only 
have to pay in-state (Idaho) tuition.
\bigskip

\item[Mouli:] Since physics has its charms, especially to young quantitatively-oriented people, did you consider 
pursuing a career in that  direction? 

\item[Jon:]  I did, but I was more enchanted with mathematics at that point, and was greatly enjoying the math courses, 
especially several courses taught by Charles Christenson.   

\item[Mouli:] Do you remember which courses, specifically? 

\item[Jon:]  The particular courses that have stuck in my mind as special were
math analysis for two semesters based on the book by Tom Apostol, set theory based on P. Halmos's book
{\sl Naive Set Theory}, and then a course on non-Euclidean geometry.
\bigskip
 \item[Richard:] How was the social life at U Idaho? Did you have lots of friends there? 
 
\item[Jon:]  There was quite an active social life at the U of Idaho, but I was rather focused on academic activities 
        during those years.  I did manage to do a bit of skiing and climbing while in Moscow, but that really increased 
substantially once I started graduate school at the U of W in 1971. 
\bigskip
\item[Mouli:] We know that you served for a while in Vietnam. Was this immediately after you finished your degree? 
       Any reminiscences of Vietnam that you would care to share with us? 

\item[Jon:] Sure! The Vietnam war was going on while I was attending the U of I, so I joined the ROTC program and 
        graduated with a commission as a Second Lt. in the Army with a two - year service obligation.  I initially
arranged a delay of service to begin graduate work, and I entered graduate studies at Yale University in September
1968.  At Yale I found it difficult to focus on studies with the service obligation looming, so I left Yale in the spring of 1969
and began Army Service in June 1969.  My first posting in the Army was to Fort Augusta, Georgia, for a Signal Officers 
Basic Course.  During that time the US first landed on the moon.   
My nine month Army service with the Signal Corps in Nha Trang 
and Long Binh, Vietnam was fairly uneventful with the exception of the occasional trip via 
helicopter to visit the signal sites on hill tops scattered over II corps (the second military region, stretching from Qui Nhon  
in the north to Phan Thiet in the south).
I remember reading Feller volume I during that time.  
The US was trying to withdraw from Vietnam, so I ended up getting a three 
month ``early out'' from the Army in March 1971. All things considered, I was very glad to 
leave the Army and Vietnam in March 1971.
I began graduate study at the University of Washington in an inter-disciplinary program in biomathematics during 
spring quarter 1971 -- starting with an undergraduate probability course taught by Albert Marshall.

\end{description}

\section{Graduate School and Rochester}  
\begin{description}
\item[Richard:] Tell us a little bit about the department 
at that time. What led you to empirical process theory and, in particular, to the topic of your PhD under 
the supervision of Galen Shorack? 

\item[Jon:]
The program in Biomath at the UW was very flexible.  The strongest sub-group 
within that program was the 
new biostat group in the School of Public Health, but as students 
we had the possibility of courses in a variety of departments, 
including the Math department.   
The probability and statistics group within the Math department had quite a strong history involving 
Bill Birnbaum, Ron Pyke, Bob Blumenthal, and Galen Shorack.  
Ron and Galen were both teaching statistics and probability during my student years.  I got interested
in the methods being developed by Galen and Ron in connection with asymptotic theory, and ended up 
doing a dissertation on some fairly technical problems connected with 
barrier crossing problems for the empirical 
d.f.  Along the way I caught the research bug:  the problems formulated at the UW 
required a number of years of 
effort to sort out, but provided ample material for further research.  
\bigskip

\item[Mouli:]
During your time at U Washington as a graduate student, did you get a chance to interact with 
       David Mason, Galen's other stellar PhD student?   Or was that later? 

\item[Jon:]  Very much so! David and I were both in the advanced probability course taught by 
Galen during either 1972-1973 or 1973-1974. John Wierman was also a fellow student in that course.   
        David did indeed also work with Galen for his PhD, finishing a couple of years later in 1977 or so.  
        Another student in the program during those years was John Crowley, a student of Norm Breslow's.  
The research by Norm and John on the large sample theory of the 
Kaplan--Meier estimator during Norm's sabbatical year in Lyon was quite intriguing, 
and was one of my motivations for learning more about weak convergence theory.
Yet another student in the program who started slightly later was Bruce Lindsay.  

\item[Mouli:] Who's John Wierman? 

\item[Jon:]  John Wierman was a fellow grad student who did a PhD with three different topics:
a Berry-Esseen theorem for U-statistics, optimal stopping, and percolation.  John has had a distinguished 
career at Johns Hopkins University, and was elected as a Fellow of the IMS in 1984.   So that probability 
class taught by Galen resulted in three Fellows of the IMS.

\bigskip
\item[Richard:] In previous conversations, you have alluded to the early pursuit of shape-restricted inference 
       in the Pacific Northwest, which was eventually to become an important area of your own research. 
There were also people in Europe looking at this in the '50s (Ulf Grenander, Constance van Eeden etc.) 
Can you tell us a little bit about these two parallel developments and also how synergies were fostered 
between these different groups? Did your early exposure to shape-restricted inference happen during 
your graduate student years? 

\item[Jon:]  Well, I was vaguely aware of work on shape restricted inference during 
my time at the UW, but it certainly did not 
       develop into a research interest until considerably later.  
       On the other hand, the contacts between the Pacific NW and 
the Netherlands was a major component of my awareness of 
international connections and activity.  Several statisticians at the University of Oregon 
(Fred Andrews, Don Truax, and (perhaps) Ted Matthes?) 
had spent sabbatical leaves at the Mathematisch Centrum in Amsterdam
during the late 1950's and early 1960's, while Willem van Zwet from the Netherlands
 had spent a leave at the University of Oregon.  Moreover, Galen, 
 who did his MS work in Math at the U of Oregon, had spent his first 
sabbatical leave in Amsterdam.  So I was well aware of the Dutch connection 
by the time I finished my PhD at the UW in 1975. 
Part of the celebration of that event was a dinner at the Space Needle 
with Galen and Frits Ruymgaart who was visiting the U of Oregon
from the NL at that time.  (Frits was, in an unofficial sense, Galen's first PhD student.) 

\item[Mouli:] Ted Matthes incidentally was my former colleague, Michael Woodroofe's advisor at Oregon! 

\bigskip

\item[Mouli:]  You joined the University of Rochester after your PhD from Washington. This is where you met one of your 
most important mentors, Jack Hall. Can you give us a brief account of your time at Rochester, and how 
Jack influenced your development as a researcher? 

\item[Jon:]  Yes, I joined the U of Rochester in September 1975, mostly because of Jack.  
Just a brief story about getting recruited there.
        When the Boeing Research Labs closed in 1970,  Al Marshall signed on for a
         one-year teaching position at the UW; 
he temporarily replaced Ron Pyke who was on a sabbatical leave during 1970-1971.  
Al then took a position at the U. of Rochester,
and he was there in the winter of 1975 when I interviewed for a job there.   He was about to leave 
Rochester and return 
to the Pacific NW and a position at the University of British Columbia.  
So I spent a few years following Al Marshall back and forth across the country. 
Joop Kemperman, a probabilist at the U of Rochester, was another important reason 
for going there.  
Jack was very positive and open about research.  
I learned a lot from him about contiguity theory, and we collaborated on 
several papers, including two papers on mean residual life and another paper on confidence bands for the 
Kaplan--Meier estimator.  
\bigskip

\item[Richard:] Tell us a bit about how the Begun, Hall, Huang and Wellner (BHHW) paper on semiparametric 
       efficiency came about. Apart from Jack, who else, if anyone, influenced your work/interests in 
semiparametric theory at that time?  

\item[Jon:]  I was trying to understand contiguity theory and how it worked during the first few years of my career. 
       Jack Hall and Bob Loynes had proved a result about the uniform integrability of likelihood ratios, and 
       that provided a starting point for reading more of the work of H\'ajek and Le Cam.  (I have called their
result ``Le Cam's fourth lemma'', but it really came from Jack Hall and Bob Loynes.)    
There was also the issue
of asymptotic efficiency of Cox's ``partial likelihood'' which was addressed 
as a special case in an interesting
 JASA paper by Efron in the late 1970's.  
 The challenge was to develop a general approach which would handle
both the Cox proportional hazards model and the models 
stemming from Charles Stein's work in the 1950's 
in which ``adaptive estimation'' was possible.  
In the late 70's and early 1980's most of the focus was on those
problems where ``adaptivity'' occurred.   
I found that several of the papers by Rudy Beran provided a readable
entry point for some of the theory current in the mid-to-late 1970's.   
I had some basic insights while preparing
for a talk at Columbia in the spring of 1980 that jump-started my own approach, 
and I pursued this during my 
initial research work at the University of Munich while on sabbatical 
leave from Rochester during 1980 - 1981.  
Although I gave an initial talk about that work at the Dutch meeting of 
Statisticians at Lunteren in November 1980,
it took 2 more years to get the paper written and on the way to publication.  \\
A side story about learning about people and their contributions:  in June of 1980, before 
starting a German language course at the Goethe Institute outside Munich, I traveled
to my first meeting in the eastern block, in Budapest, Hungary.  During that meeting
Willem van Zwet was going to have dinner one evening with a young Russian fellow 
by the name of Boris Levit, and he invited me to join them.  We had a very pleasant dinner
in a castle above the river in Budapest, and it was clear that Willem was trying to figure out
how to get Boris out of Moscow and into a position in the Netherlands or somewhere else in the
west.  At that time I did not have a clue about the scientific background of Boris 
or his accomplishments.   Quite by chance while browsing in the math library in Munich
later that Fall I ran across a paper by Boris, and began to make the connections:
it turned out that he was doing (together with Yu A. Koshevnik and others in Russia) the same kind of thing that 
I was trying to do in connection with the BHHW paper, but from a different (nonparametric) perspective.   
\bigskip

\item[Mouli:] The BHHW (1983) paper had a strong impact on the subsequent development of semiparametric theory 
        in the 1980's and 1990's. How would you assess the legacy of that paper?   

\item[Jon:]  The (1983) BHHW paper certainly got some of the simpler, broad brush, parts of the theory right, but there were many 
        subtleties that were glossed over or even given a somewhat mis-leading treatment.  My only excuse is 
that I was struggling to learn enough math to formulate the problems correctly.  In another respect,
the paper might have had a bigger impact if we had gotten the title right by somehow including the word ``semiparametric".
Many of the subtleties became more apparent during the work on BKRW (1993) over the next 10+ years 
with Peter Bickel, Chris Klaassen, and Ya'acov Ritov.  In any case the main thrust of the BHHW paper was 
on target, and did have the effect of moving the focus away from the special cases involving ``adaptive estimation".
\bigskip

\item[Richard:] You spent your first sabbatical at the University of Munich under the auspices of a Humboldt Foundation 
       Grant in 1980-81. Can you tell us a bit about that experience? Did you, in particular, get to interact there with 
some people whose academic ideas were influential in the future? Did you meet Peter Gaenssler, whom you 
list as one of your mentors, at that time? 
\item[Jon:] Yes, the choice of Munich resulted from a correction to one of my early papers pointed out by 
Peter Gaenssler and Winfried Stute.  Peter nominated me for the Humboldt Fellowship, and the 
Humboldt funding resulted in the year
in Munich.  The group there had a seminar going on martingale theory, and that was interesting to me because 
of the work on the Kaplan--Meier estimator via martingale theory by Richard Gill and the Copenhagen school.
During the year in Munich I made visits to the Netherlands at least twice:  during the Fall of 1980 Richard was
away in Copenhagen working with Nils Keiding and  Per Kragh Andersen, but I did succeed in tracking him down 
in Amsterdam during the Spring of 1981.   I should note that Peter Gaenssler made very effective use of the
Humboldt Fellowships over the years, supporting not only myself in this way, but also David Pollard 
(before me) in Bochum 
before Peter moved to Munich, and then David Mason a few years later (also in Munich).
\bigskip
\item[Mouli:] Any other specific memories of Rochester? Michael Akritas once told me that he, you and Jack 
 used to go skiing quite a bit at Rochester!  
\item[Jon:] Yes!  The Wednesday night ski trips to Bristol Mountain south of Rochester were a regular event!  
Bristol had a thousand feet of vertical drop for skiing, and it had plentiful artificial snow during the 
cold Rochester winters, so I managed to do a fair amount of downhill skiing. 
\item[Mouli:] I thought Rochester gets a lot of natural snow as it is...they still needed artificial snow? 
\item[Jon:] But at a ski area the snow gets pushed around and often forms bumps or moguls.  Since it was often cold 
enough to make snow, the local ski area took advantage of every opportunity to make more.

\begin{figure}[ht!]
\centering
\includegraphics[width=65mm,height=70mm]{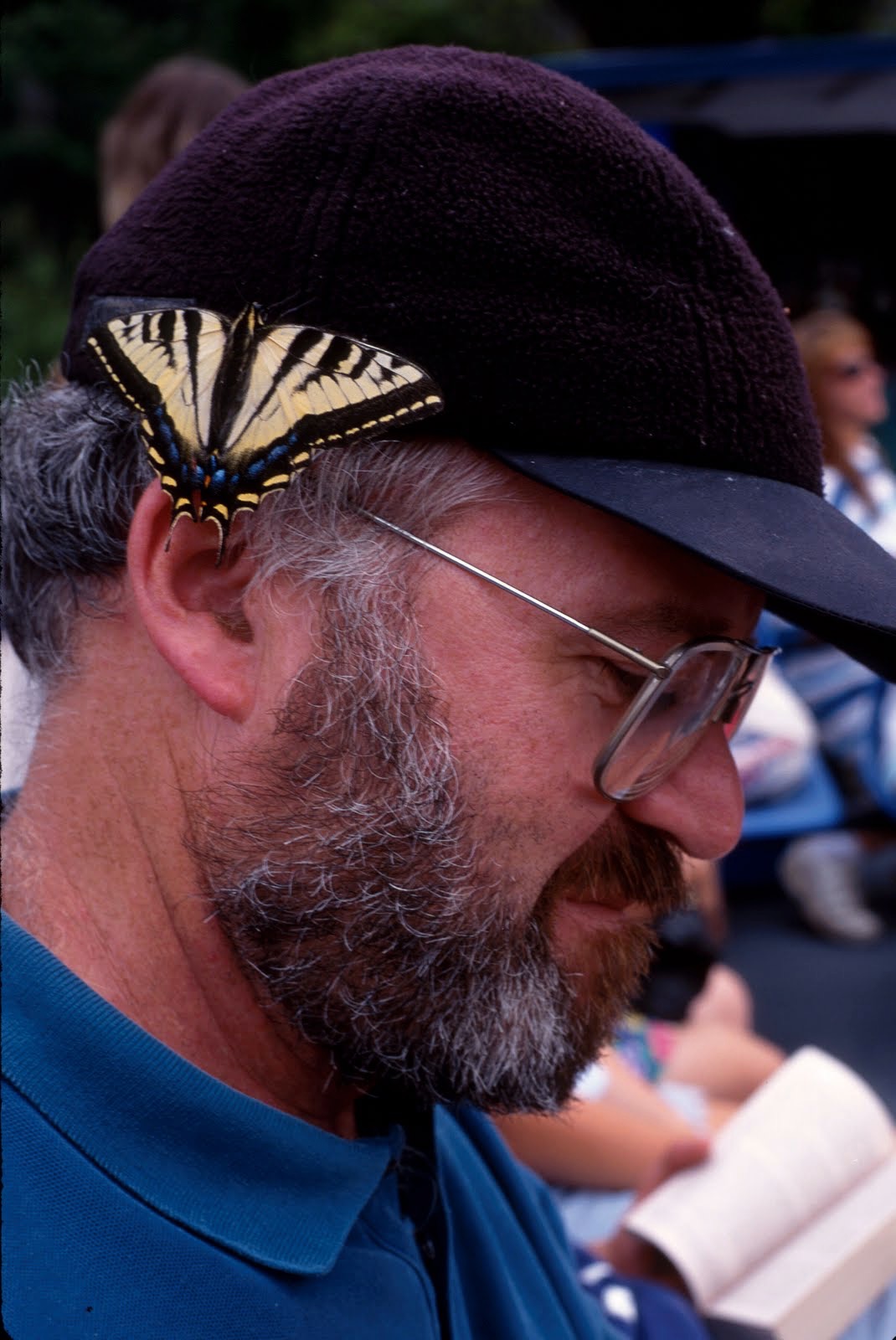}
\caption{A butterfly seeks the nectar of mathematical statistics.} 
\label{fig:figure2}
\end{figure}

\end{description}

\section{1983 Onwards: University of Washington}
\begin{description}
\item[Richard:] I suppose one could say 1983 defines the beginning of a new phase of your academic career as you 
       returned to the University of Washington and have stayed there ever since. What prompted you to move 
back to your alma mater? Did the appeal of the Northwest play a big role in this? 
\item[Jon:]  The appeal of the Pacific Northwest played a huge role.  Proximity to family was another major motivating force.
By this time my father and younger brother had moved to Moscow, Idaho, for my father's retirement and 
 his ``second career'' on natural areas.   The creation of the Statistics Department at the U of Washington in 1979 
also played a huge role.  During the time at Rochester I visited the UW during the summers of 1977, 1979, and 1982.
Those visits kept the UW connections going during the Rochester years.
\bigskip
\item[Mouli:] Your first book on Empirical Processes with Galen Shorack was completed in 1986, a relatively short while 
       after you moved back to Washington. When did the two of you decide to go ahead with this project?
       How did it develop? 
\item[Jon:] Galen invited me to join him in writing the book toward the end of my summer visit to the UW in 1977.  Our work on 
the book really got going in 1978-1979,  but then it slowed down when I was on sabbatical leave in 1980-1981.  
In any case, the whole project lasted for nearly 9 years, 1977 - 1986.  
Our editor at Wiley, Bea Shube, was very supportive and kept urging us on.
\item[Richard:] Given the length of the book, that's still a pretty decent rate of writing!

\bigskip

\item[Richard:] It is interesting to note that your four books were published within a span of 10 years: your book with Galen 
       came out in '86, the one with Aad van der Vaart in '96, and your other two books in '92 and '93.  What motivated 
this very active book-writing phase of your career and how do you see this in hindsight? \\
\item[Jon:]  Well, that is difficult to explain.  
 In part it resulted from a desire on my part to understand the current state of 
        limit theory in statistics and some of the unifying tools.  
 As the process of doing research developed, it was fairly clear that there were 
 big gaps in the systematic coverage of various areas, including empirical process theory.  
 And it was also clear that  general empirical process theory -- which was developing rapidly 
 during the mid-1980s (thanks to the pioneering 
 work of Dick Dudley, David Pollard, Ron Pyke, Evarist Gin\'e, Joel Zinn,
 Mike Marcus, J{\o}rgen Hoffmann-J{\o}rgensen and others)  would be extremely useful for all sorts of statistical questions.  
Fortunately, I managed to find excellent co-authors who were also interested in some of 
these developments.  

\bigskip 

\item[Mouli:] Tell us a bit about your famous book with Bickel, Ritov and Klaassen on semiparametric theory. 
       How did the synergies that led to this book take shape? You've of course written papers with all three of them, 
       but how did that book develop?   \\
\item[Jon:] The BKRW book resulted from an invitation that Peter Bickel received from Bob Serfling at Johns Hopkins 
University to give a series of lectures on estimation theory in the summer of 1983.  
Peter initially invited Chris Klaassen
and myself to join him in writing up lecture notes on semiparametric theory based on his 
course at Hopkins.  
Shortly thereafter Ya'acov Ritov began a series of visits to Peter, and they began solving a 
set of basic problems 
connected with the theory, so Ya'acov joined the project as well.  During my second 
sabbatical at Leiden in 1987 - 1988,
Aad van der Vaart was just finishing his PhD work with Klaassen and van Zwet on semiparametric 
theory, so Aad and 
I had many fruitful discussions about the theory during the Fall of 1987, and some of the examples 
eventually found their
way into Aad's very nice 1991 Annals paper.   
A number of papers were generated during that time by just trying to figure out how the theory interacted with several basic examples. 
\bigskip
\item[Richard:] Before we get to your next two books, let's talk about the two scholars with whom you wrote 
       books, and with whom you have had very long term interactions: 
       Piet Groeneboom and Aad van der Vaart, 
and also about your general connections to the Dutch statistical school. 
Could you elaborate on how your interactions with Piet and Aad, and in general, the Dutch school evolved? 
\item[Jon:]  As I have mentioned above, my awareness of the Dutch school of statistics and probability started developing
       during my time as a graduate student in the early to mid-1970's. I met Aad in Amsterdam at the 1985 ISI Meeting.  
       Aad had started his PhD work in Leiden with Willem, and Willem introduced me to Aad at that meeting.  
       The meeting in Amsterdam was followed by a satellite meeting 
in Maastricht.  I remember doing several walks around Maastricht with Aad and Marie Huskova from Prague, 
during and after that satellite meeting.  

Because Piet was in Seattle and Berkeley while I was in Rochester and Munich, my memory is that 
I did not meet him until 1987 or so, when I was on sabbatical leave in Leiden.  

Soon after going to Rochester in 1975 I attended
the ``Purdue Symposium" on statistics in the spring of 1975.  Willem van Zwet was also attending the 
meeting, and we ended up sitting together for the conference dinner.  It happened that 
Willem was very much aware of some of the results in my PhD thesis, 
but he felt that I had tackled the wrong problem in some sense:  
I had proved a law of the iterated logarithm for linear combinations of order statistics, but that we 
did not yet have a good strong law of large numbers for such statistics.   
That suggestion took root with me and 
I went to Seattle for a few weeks and spent much of the time 
working out my version of such a theorem.  
The result was published in the {\sl Annals of Statistics} in 1977;  
Willem's much more elegant and general paper
on the same topic was published a year later in the {\sl Annals of Probability}.  
In any case, my connection with Willem 
has been a very important component of my relationship with the 
whole group of statisticians and probabilists in 
the Netherlands, including Frits Ruymgaart,  Chris Klaassen, Richard Gill, 
Piet Groeneboom, Geurt Jongbloed, 
and Aad van der Vaart.  It has been a great honor to collaborate with all of them.   
In the end, the largest part of 
my collaborations have been with co-authors in 
the Netherlands.  It has been a great experience! 
\bigskip
\item[Mouli:] Was it primarily Piet who got you interested in the field of shape-constrained inference, the topic of this 
       special issue, and an area that has been a core theme of much of your research in the second half of 
your career?  

\item[Jon:]  Yes!  I learned about Piet's 1989 Rollo Davidson Prize paper in the mid-1980's 
 long before it was finally published in (1989).  This paper remains a tour-de-force benchmark in terms 
 in the whole area.  Piet's paper provides a complete and detailed description of Chernoff's limiting distribution of the 
 location of the maximum of two-sided Brownian motion minus a parabola.  
 It illustrates the great value of focussing on a concrete problem.  
 Piet has returned to this theme in the last few years, giving new proofs of the results in his '89 paper
in collaboration with Steve Lalley and Nico Temme.   The area of shape constraints is building up a store-house 
of further problems in this direction which await either the further  research of Piet himself or the 
interest of future research workers.  

\begin{figure}[ht!]
\centering
\includegraphics[width=75mm, height=60mm]{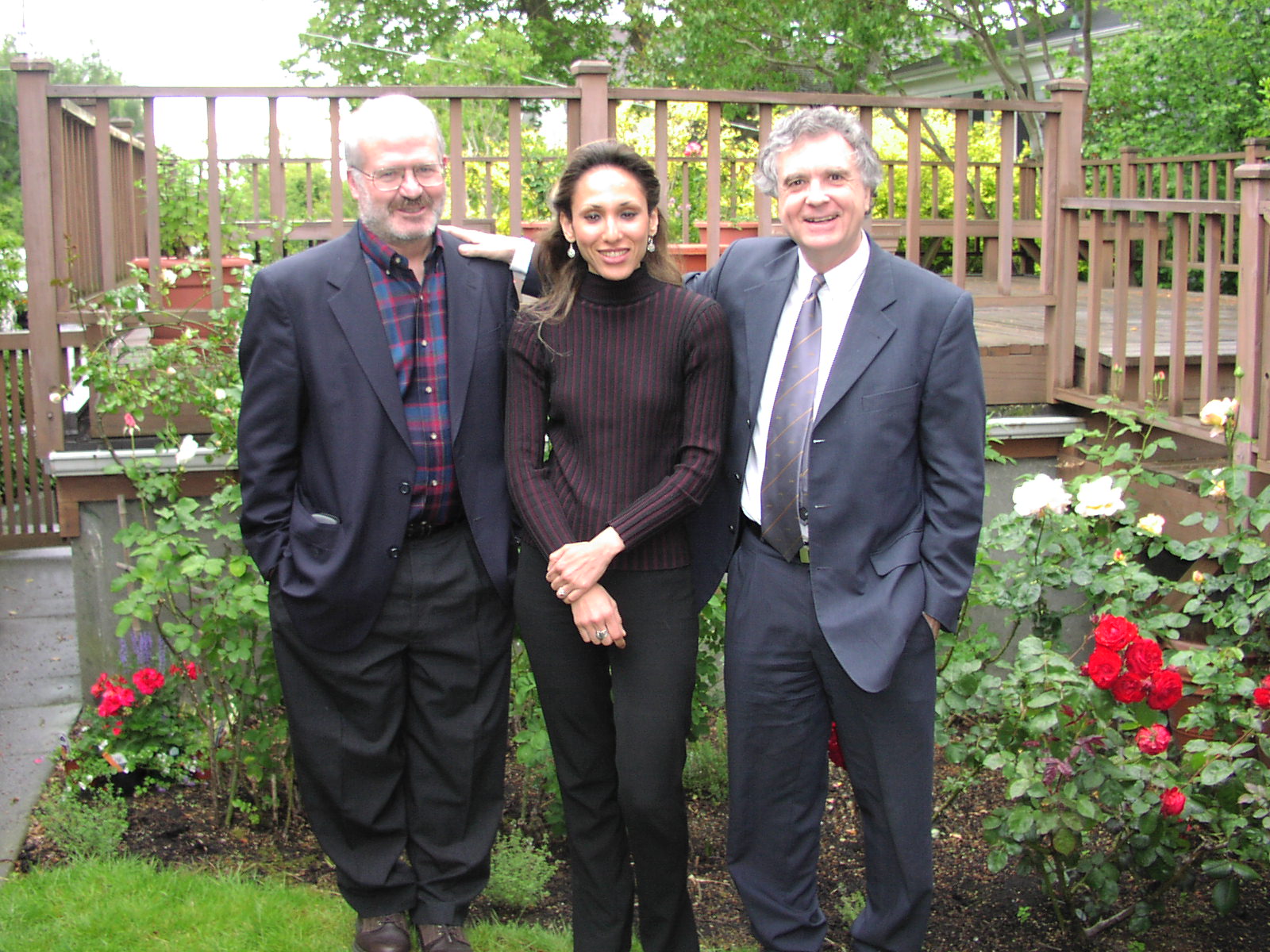}\includegraphics[width=75mm,height=60mm]{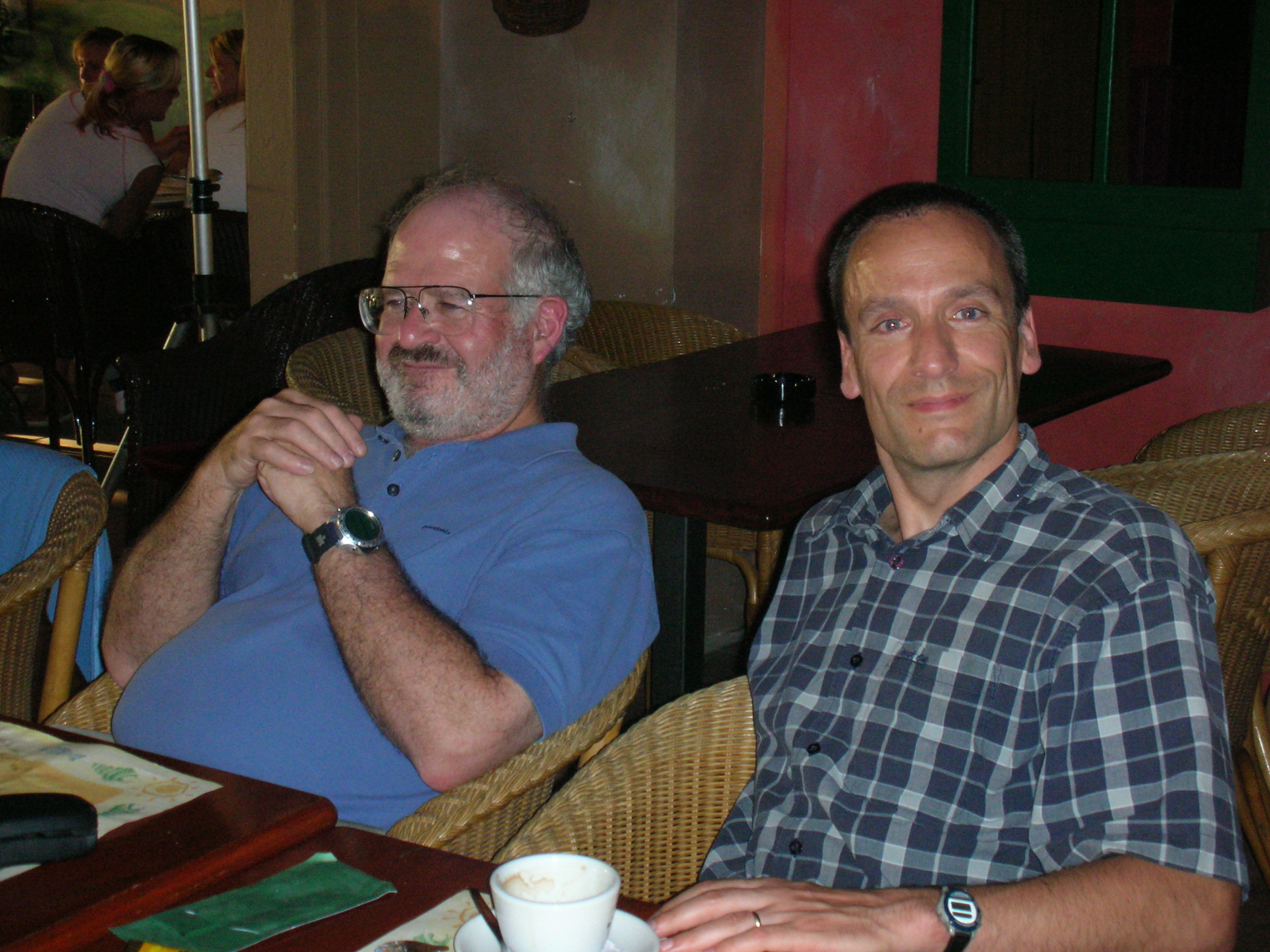}
\caption{Left: Jon, Fadoua Balabdaoui, and Piet pose amid verdant surroundings in Seattle 2004; 
right: Jon wears a smile of infinite contentment, and Lutz strikes a smart pose for the camera 
(taken during a dinner at Noordwijk during Piet's 65th Birthday Meeting in 2006).} 
\label{fig:figure2}
\end{figure}

\item[Richard:]
Tell us about your book on Information Bounds and Nonparametric Maximum Likelihood Estimation 
        that was published by Birkhauser in 1992.  What prompted the writing of that book? The first part of the book 
deals with semiparametric theory, which is also covered by BKRW, the second part is largely on interval-censoring 
models (current status data and Case 2 interval censoring). Would it be correct to suppose that your evolving 
interests in interval censoring at that time had something to do with this book? You subsequently wrote a number of 
interesting papers on the current status model in particular. 

\item[Jon:] 
Piet was invited to give a series of lectures at G\"unzburg in southern Germany in 1991 or 1992.  
Piet had been working on the theory of estimating a distribution function with interval censored 
data during 1987, and we discussed that work at Lunteren in November 1987.  
I spent time trying to work this model into the book with Bickel, Klaassen, and Ritov
during the spring of 1988 and I had a number of exchanges with Piet about all that in 1988 - 1989.  
So Piet invited me to join with him in writing up some notes from his lectures.  I ended up learning a lot about 
        Piet's methods and approaches during that time, and that was my start on serious involvement with shape-constraints.
I was fascinated with the fact that the same limiting distribution (non-standard; Chernoff's distribution) 
was arising in at least two quite different nonparametric monotone function estimation problems.  
It is now well-known that it arises
in a large class of such problems, but for me in the early 90's this was new and interesting.

\item[Mouli:] It seems the Chernoff limit result for Case 1 censoring appears for the first time in this book? It was never 
published in a journal, was it? 

\item[Jon:]  That is correct as far as I know.  Piet had written a technical report at the University of Amsterdam 
in 1987 that contained the result.  
\bigskip

\begin{figure}[ht!]
\centering
\includegraphics[width=110mm]{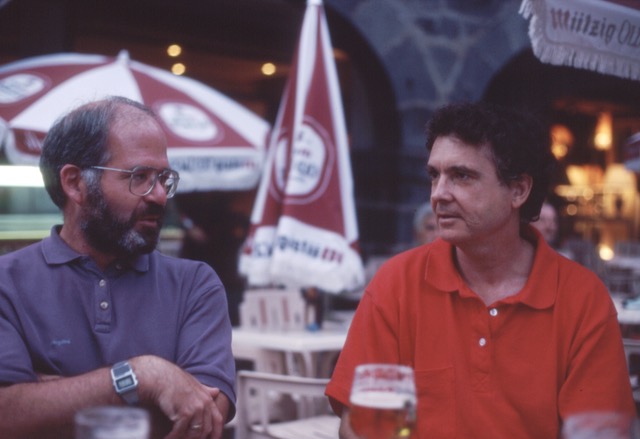}
\caption{Jon manages to distract Piet from his beer at  St. Flour, 1994}
\label{fig:figure3}
\end{figure}

\item[Mouli:] Let's talk now about your most highly cited book (with Aad), which as of going to press, has garnered 6000 citations. 
This book, by and large, introduced the general statistical audience to the tools of modern 
       empirical process theory, going beyond the more traditional empirical process theory covered in your earlier book, 
       and clearly fulfilled a dire need. It would probably not be amiss to say that this book is a standard toolkit of a large 
       majority of mathematical statisticians and theoretically-inclined methodologists today. 
       How did this book come about? 
       And having witnessed its grand success, how do you feel with hindsight? 

\begin{figure}[ht!]
\centering
\includegraphics[width=110mm]{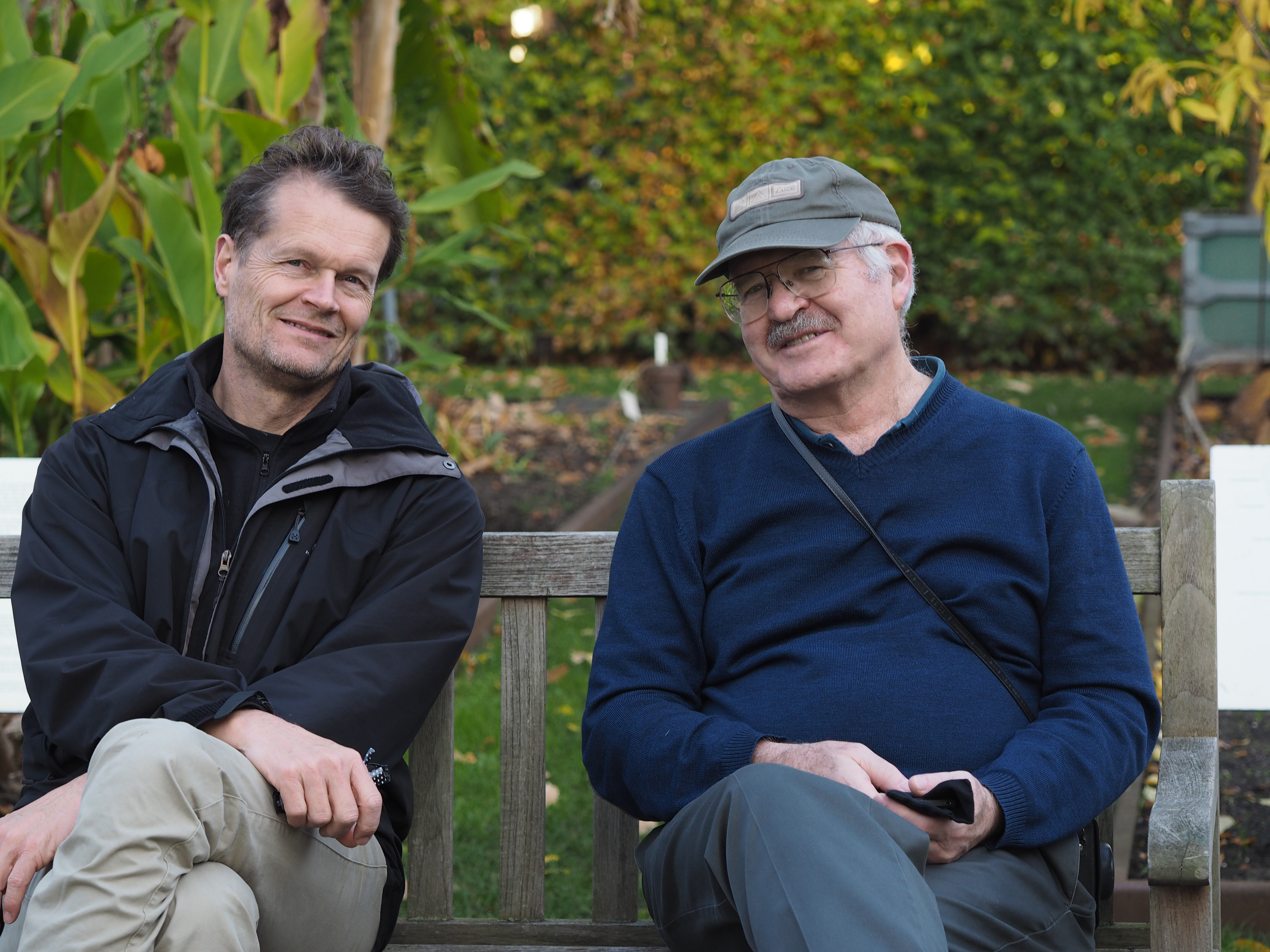}
\caption{Aad and Jon relax at the Hortus Botanicus, Leiden 2015}
\label{fig:figure4}
\end{figure}       

\item [Jon:]  The success of the book with Aad has been very gratifying.  I learned an enormous amount from 
working on the book and from working with Aad.  
We started by trying to write down some of the basic theory needed to develop several examples.   
One particular motivating problem was to justify the approach that I had suggested in my 1989 discussion 
of an important paper by Richard Gill,  
``Non- and semiparametric maximum likelihood estimators and the von Mises method (Part I)''.  
Scand. J. Statist. 16 (1989), 97 - 128.
The issue was justification of Hadamard (or compact) differentiability in a general setting.  
To do that Aad and I needed 
a generalization of Herman Rubin's  generalized Mann - Wald (or continuous mapping) 
theorem involving a sequence of 
functions $\{ g_n \}$ rather than just one fixed continuous function $g$, to the Hoffmann-J{\o}rgensen 
weak-convergence framework.  This is just one example of problems in which we needed extensions 
of the classical weak convergence theory.   And in the end the simplest approach was to write a 
book where all the needed theory could be collected in one place.
[We also wrote a paper on our early efforts to understand the H-J theory, 
``Prohorov and continuous mapping theorems in the Hoffmann-J{\o}rgensen weak convergence theory, 
with applications to convolutions and asymptotic minimax theorems'',
which received rather negative reviews from the journal where we submitted it at the time.  
It seems that most of the reviewers of that paper felt that the heavy theory was simply over-kill.  
That might have been true, but I would argue that it was very useful theory in the longer term.  
For example, it has recently been useful in the study of multivariate distributions via optimal transport 
theory; see e.g. Chernozhukov, Galichon, Hallin, and Henry (2017).]

We had the good fortune of producing the book at an opportune time -- when the theory 
had developed to a point where 
quite a bit was known, but there also remained quite a few open problems.   
When we wrote the book the theory of 
concentration inequalities was still under very active development.  There are several 
other instances of areas that
were still developing quite quickly, and which did not make it into the 1996 book. 

\item[Mouli:] The recent Boucheron, Lugosi, Massart book on Concentration Inequalities probably 
fulfills the same sort of need that your book on Empirical Processes did at that time! 

\item[Jon:] Yes, I agree.  
Of course the book by Boucheron, Lugosi, and Massart focuses on 
inequalities per se, while the 1996 book with Aad covers other ground as well.  
The success of the Boucheron, Lugosi, and Massart  book is ample testimony to 
the importance of inequalities!

\bigskip

\item[Richard] Is a revision of {\sl Weak Convergence and Empirical Processes}
 forthcoming sometime in the near future? 
\item[Jon:] Yes,  at least we hope so!  Aad and I sent a revision to Springer in October last year.  
Unfortunately we have not received a definite response from Springer yet.  I continue to 
hope that Springer will publish a revision within the next year.
[The revision does include quite a few improvements and additions, so it will be well 
worth buying a copy when it does come out!]
\item[Richard:] That's excellent news!  
\item[Mouli:] Count me in! I already have two copies, one for the home, the other for the office...
and maybe the new edition I'll put in my travel bag!! Ha, ha! 

\bigskip

\item[Mouli:] You had an active collaboration for many years with Norm Breslow at Washington, 
who was also a close friend. 
Presumably, this was also influential in triggering your interests in models with Biostatistical applications. Would 
you tell us a bit about your work with Norm?  
\item[Jon:] 
  As I mentioned above, my first interactions with Norm came indirectly via John Crowley and their joint work on 
         the large sample theory of the Kaplan-Meier estimator.    When I got back to 
         Seattle and the UW, I gradually got connected with a group of 
         ski-mountaineering folks through Norm and others
in the Math Department.  And then Norm organized trips to Nepal in 1989 and 1996.  
In between Norm and I did quite a lot of climbing together - - including successful climbs 
of Sloan (1987), Columbia (1992), Formidable (1995), Fernow (1997),  
and unsuccessful attempts to climb Jack Mountain (1993, just as his first grand-child was about to be born),  
Bonanza Peak, and Dumbell.  
We also joined others in about 8 or 9 spring ski-mountaineering trips to the British Columbia 
Coast Range, starting in the early 1990's and lasting until about 2013.

Our collaborative scientific work only began in the mid 1990's 
when Norm got me interested in two phase designs and the resulting statistical issues arising in connection 
with these designs and semiparametric models.   Between 1995 and 2015 we wrote 5 or 6 
papers on this topic, and my 
interest in this area has continued since Norm's death in 2015.  
\bigskip

\item[Richard:] The theme for your 65th birthday celebrations in Seattle in 2010 involved the term 
        `From Probability to Statistics and Back'. 
        You have mentioned to us on certain occasions how you have always 
        enjoyed working at the interface of probability and statistics, being able to 
        drift from one to the other and back. 
To what extent has this informed your research? Do you feel that there is not that much of this happening 
anymore, now that probability and statistics have diverged somewhat, as disciplines? 

\item[Jon:] 
The freedom to go back and forth between parts of probability theory and statistics has been very important
          for me.  It seems that these connections are less important for many people working in only one or the other
of these areas, but I have enjoyed being able to spend time learning different bits of probability theory and using
them to help address statistical problems.  This type of activity is still going on, but perhaps at a reduced level 
and in somewhat different directions than when I started my career.   
\bigskip

\item[Mouli:]
You have had numerous students over the years working on the different areas of your interest, more 
         than 30 counting the ones who are still working with you. And much of your core work is contained in their 
         dissertations. Tell us about your student-advising experience a bit, and how you have found it rewarding. 

\item[Jon:]  Supervising PhD students has been both rewarding and challenging.  Every student is different,
so the trick is to try to find the right match between the student and the problem(s).  Many students already know 
more or less what they want to do and are very capable.  I  have been very fortunate in having supervised 
a number of very strong and creative students.  And then it is often the case that I end up learning more 
from them than they learn from me!  

\item[Richard:] I agree completely!

\bigskip

\item[Richard:] How about post-doctoral supervision? 

\item[Jon:]I have had very little grant support available for postdocs over the years, and hence 
         I have only supervised two post-doctoral students:  
         Hanna Jankowski (from Toronto), and Adrien Saumard (from Rennes and Pascal Massart's group in Paris).
Both of those experiences were very positive and enjoyable from my point of view.  
Hanna and I studied estimation of 
convex hazard functions and estimation of discrete monotone distributions (resulting in 4 joint papers).  
During his one year stay at the UW 
I managed to get Adrien interested in Efron's monotonicity theorem and the whole area of log-concavity.
We have written three papers together so far, and at least one more paper is in the works.    
\bigskip

\item[Mouli:] Would you like to mention some of your own favorite papers? Maybe the top five in your view? 

\item[Jon:]  Sure!  My favorite 5+ papers include: 
\begin{itemize}
         \item my 1978 ZfW paper on ``ratio limit theorems''.  (I believe that Jack Kiefer was the AE for ZfW 
                at the time and handled this paper.)
         \item the BHHW 1983 {\sl Annals} paper mentioned earlier; 
         \item the 1988 {\sl Annals} paper with Richard Gill and Ya'acov Ritov on biased sampling;  
         \item the 1993 {\sl Ann. Prob.} paper with Jens Praestgaard on bootstrapping with exchangeable weights;
         \item  the 2008 {\sl Annals} paper with Leah Jager on goodness of fit tests based on R\'enyi divergences   
         \end{itemize}
 Several of the papers on shape-constrained estimation should also be listed here, but we can return to those later.
 \bigskip
 
 \item[Richard:]  Apart from Jack Hall and Peter Gaenssler, would you like to tell us about some of your other major academic 
         influences that have not been covered in our conversation thus far? Maybe Evarist Gin\'e, Richard Gill, Lutz D\"umbgen?  
\item[Jon:]    
Richard was very influential in getting me going on martingale theory in connection with survival 
analysis.  His work on the Cox model with Per Kragh Andersen played a big role in getting 
solid large sample theory settled for the Cox model.   
Richard's 1989 paper on 
``Non- and semiparametric maximum likelihood estimators and the von Mises method'' convinced 
me of the importance and utility of Hadamard differentiation.  
(I have written two or three papers 
with Richard over the years, and have greatly enjoyed that collaboration.)

(The late) Evarist Gin\'e and Joel Zinn were two very active participants in the ``High - Dimensional Probability'' 
group starting back in the 1970's.   I started attending the meetings of this group in the early 1990's 
and organized one of the meetings, HDP II, in Seattle in 1999.

The papers by Gin\'e and Zinn 
on general empirical process theory in the 1980's and early 1990's were inspirational
in terms of their scope and generality.  Evarist became a great friend after a visit to Storrs in 1996.
We only wrote two papers together, but it was always a great pleasure to meet up with him at the HDP
meetings or elsewhere and talk about empirical process theory.   He passed away far too soon.   

Lutz D\"umbgen and I have had a lot of fun working on several different problems, 
including a great collaboration with Sara van de Geer and Mark Veraar 
to better understand Nemirovski's inequality, 
proving a version of Marshall's lemma for convex density estimation, 
proving a neat law of the iterated logarithm for Grenander's estimator, and ... figuring out new 
nonparametric confidence bands for distribution functions related to an intriguing test statistic due 
to Berk and Jones.  
That last project is still ``under revision'', but it will be a very nice paper when it is finished!
Lutz introduced both Vera and me to swimming down the Aare in Bern.  I am pretty sure that 
Vera would not do that again, but I might be up for another go at it --  in the summer when the 
river has warmed up!

\item[Mouli:] I remember helping out with the registration desk at the 1999 HDP II meetings in Seattle, as a graduate student! 

\end{description}
\begin{figure}[ht!]
\centering
\includegraphics[width=110mm]{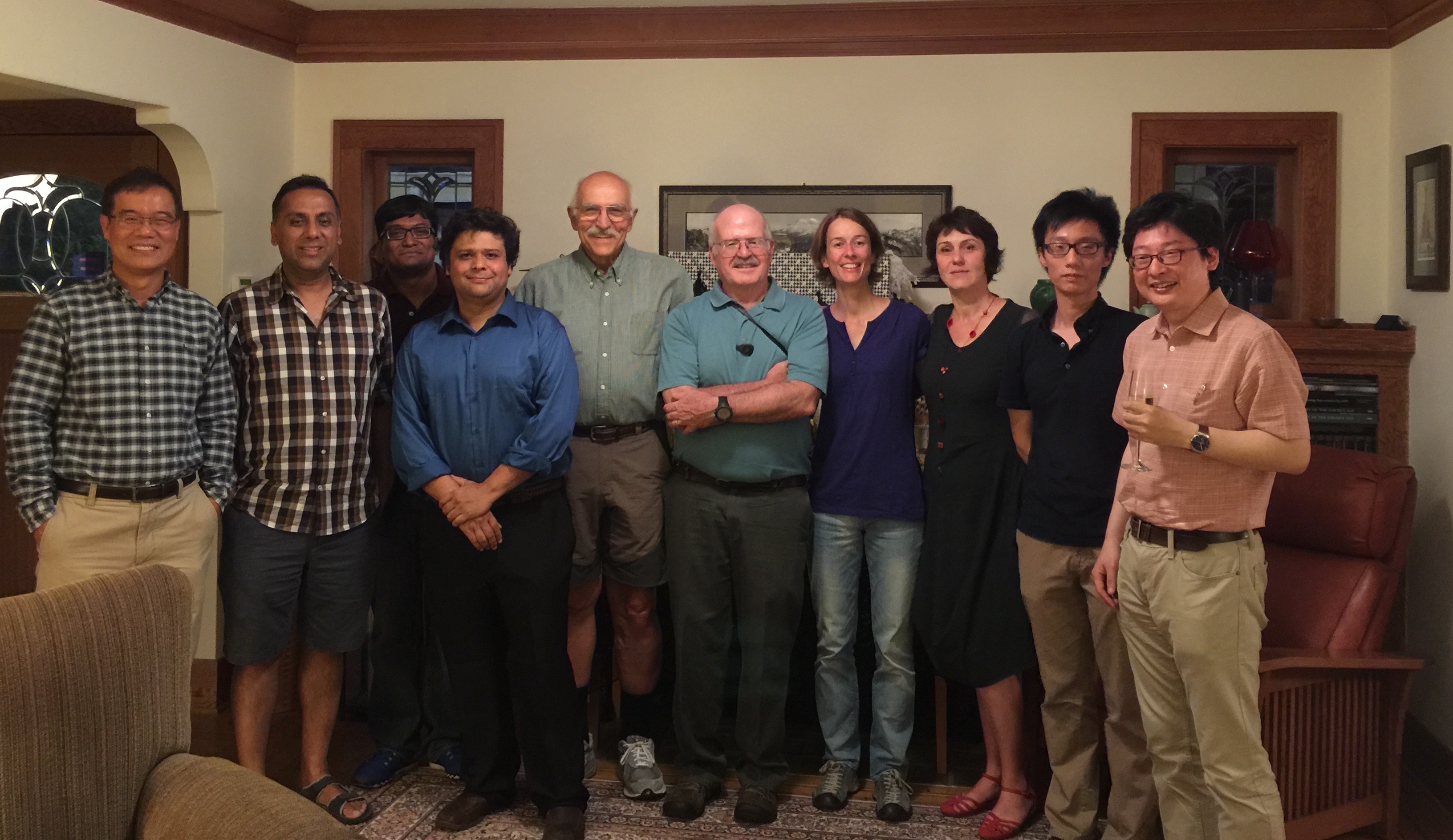}
\caption{A glimpse of (part of) an academic family: from left to right are Bin Nan (S) , 
Nilanjan Chatterjee (S), Bodhisattva Sen (Jon's academic grandson), 
Mouli Banerjee (S), Galen Shorack (Jon's advisor), Marloes Maathuis (S), 
Florentina Bunea (S), Roy Han (S), Takemi Saegusa (S). (`S' means Jon's student)}
\label{fig:figure5}
\end{figure}

\section{Shape-Constrained-Inference} 
\begin{description}
\item[Mouli:] Let's talk a bit more about shape-restricted inference, since it has been 
a major theme of the later part of your career.  What would you say most drew you to the area of shape-constrained inference? 
What is it about the area that you particularly like? 

\item[Jon:]  I have been attracted to shape-constrained inference by 
the nonstandard nature of the limit theory together with the large number of open
problems.  The strong cross-connections with inequalities and convex analysis 
is another attractive feature in my view.  There are further cross-connections with 
probability theory via the non-standard limit theory, which we still understand 
only partially.  Piet Groeenboom's 1989 
ZfW\footnote{ZfW=Zeitschrift f\"ur Wahrscheinlichkeitstheorie und Verwandte Gebiete, now known 
as PTRF = Probability Theory and Related Fields} 
paper illustrates the possibilities 
in this direction.   It is a very rich area with lots of opportunities 
for both application and further theory.  

Another attraction has been the personal connection to particular people
involved in developing the theory.  For example, in 1976 or 1977, I attended a regional
meeting at Cornell on {\sl Stochastic Processes and Applications}, with some sessions
on statistics as well.  I remember Jack Kiefer giving a talk about his 1976 
ZfW paper 
with Jack Wolfowitz on what came to be known more generally as ``Kiefer--Wolfowitz theorems''.

Let $\mathbb{F}_n$ denote the usual empirical d.f. and let $\widehat{F}_n$ denote the maximum likelihood
estimator of a concave distribution function (i.e. the distribution function corresponding to the 
Grenander estimator of a monotone density).  By building on a key lemma due to Al Marshall (appropriately 
enough known as Marshall's lemma), Kiefer and Wolfowitz (1976)  proved, with $\| \cdot \|$ denoting the 
supremum norm,  that 
$\| \FF_n - \widehat{F}_n \| = O( (n^{-1} \log n)^{2/3} )$ almost surely under curvature hypotheses 
on the true d.f. $F$.   
When Lutz D\"umbgen and Kaspar Rufibach and I proved an analogue of Marshall's lemma 
for convex decreasing densities in 2007 (just in time for Piet's 65th birthday conference in Leiden),
I knew that it should be possible to prove at least a partial analogue of the Kiefer--Wolfowitz theorem
in this case for the least squares estimator.  
Fadoua Balabdaoui and I managed to accomplish that in a paper that appeared 
in the same Festschrift volume for Piet.  The story is not over, though.  These results remain incomplete
and somewhat unsatisfactory:  we still lack a good Kiefer--Wolfowitz theorem for the Maximum
Likelihood Estimator of a decreasing convex density, 
and we also still lack a 
Kiefer--Wolfowitz theorem for the log-concave MLE on the line, 
much less for the MLE of a log-concave density on $\RR^d$.

\item[Richard:] Yes, recently Arlene Kim, Aditya Guntuboyina and I managed to 
prove a version of Marshall's inequality for univariate log-concave density estimation, 
but as you say we don't have a Kiefer--Wolfowitz theorem.  

\bigskip
\item[Richard:] How many of your students and post-docs have worked on this area? 
\par\noindent
\item[Jon:]  To date 9 of my 29 past PhD students and both of my current PhD
students have worked in this area for a total of 11 PhD's.  This represents a bit more than 
one third of my past and present PhD students.  And, of course, both of my post-docs! 
\bigskip
\item[Mouli:] What do you think has spurred the surge in interest in shape-constrained inference over the last decade? 
       Do you think the close connections to convex optimization, and more broadly, convex geometry have helped 
       get a broader audience interested? 
\item[Jon:] Yes, the connections to convex optimization, convex geometry, and convexity based inequalities 
have helped to spur the increasing interest in shape-constrained approaches.   I am still working toward a better
understanding of the collection of inequalities connected with the Brunn--Minkowski theory as outlined in 
the wonderful survey paper by  R. J. Gardner (2002). 
\bigskip
\item[Richard:] 
What do you see as the next main challenges of the shape-constrained community?   
\item[Jon:] 
\#1 There are many challenges arising from the development of shape-constrained 
procedures in higher dimensional settings.  This is true both for convexity constrained 
estimates of regression functions and for convexity constrained density estimation.  
At the present time we lack answers to many fundamental questions about these estimators,
not to mention inference procedures beyond estimation. \\ 
\#2 A further challenge is to create new shape-constrained models which do not break down 
in high-dimensional settings.  \\
\#3  Developing methods for shape constraints in connection with semiparametric models
of interest in applications.
\bigskip
\item[Mouli:] Is the shape-constrained community having as much impact on applications as it should? 

\item[Jon:] No, probably not yet.  This is partially due to the difficulty of the theory and lack of 
readable expository and review material in the area.
The recent book by Piet and Geurt  
{\sl Nonparametric Estimation under Shape Constraints: Estimators, Algorithms and Asymptotics},
 might provide some push for changing that.  
 Development of faster computational methods might well play an important part in increasing the 
 number of applications, but the 
 community also needs to do more work to provide inferential methods beyond estimation.  
\bigskip

\item[Richard:]
Of the results in shape-constrained inference that you have established, which ones are your favorites?  
\item[Jon:]  Among my favorites are:
\begin{itemize}
\item the 2001a,b {\sl Annals} papers with Piet and Geurt;
\item  the 2001 {\sl Annals} paper with Mouli; 
\item  the 2009 {\sl Annals} paper with Kaspar and Fadoua on the pointwise limit theory 
for log-concave maximum likelihood estimators on $\RR$.  
\end{itemize} 
I especially like the limit theorem for the 
log-concave mode estimator in the latter paper!  

\bigskip
\end{description}

\begin{figure}[ht!]
\centering
\includegraphics[width=110mm]{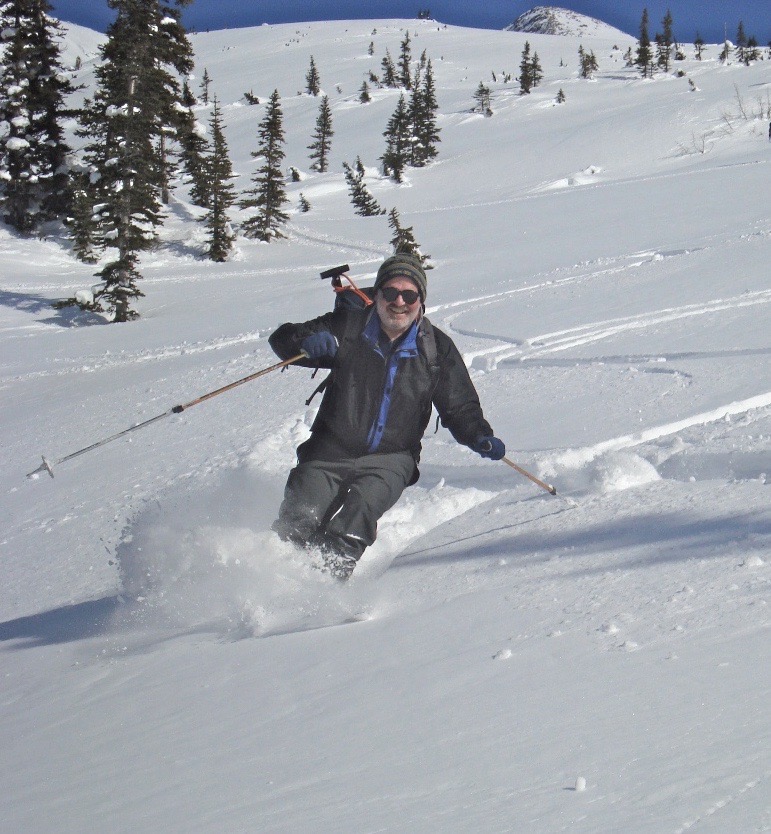}
\caption{Jon traces out complex shapes in the snow that will require formidable future analyses.}
 \label{fig:figure6}
\end{figure}

\section{The Broader Profession and Honors}  
\begin{description}
\item[Mouli:] You have been involved in service to the broader profession at several levels. 
In particular, you served as Co-Editor of the 
Annals of Statistics with John Marden from 2001-2003. Would you tell us something about your experience?

\item[Jon:] Editing the Annals was a marvelous educational experience.  
It was more work than I had anticipated, but it provided
       a wonderful overview of the breadth and depth of the current 
       research interests of people all over the world at that time.
       It was also quite a broadening experience in terms of working with a 
       fairly large group of excellent people as Associate Editors.  The other two journals 
       I edited for the IMS,  {\sl Statistics Surveys} and {\sl Statistical Science}, were also
      educational and broadening, but in very different ways than the {\sl Annals} experience:  
      they are simply very different journals.
      {\sl Statistics Surveys} was just getting underway at that point, and in my view is still 
      underused and probably under-rated.   The probability side of our community has been ahead 
      of the statistics community in terms of making good use of 
      their version, {\sl Probability Surveys}, as well as being ahead in terms of electronic venues for publication
      more generally.
 \bigskip
 \item[Richard:] How, in your view, has the Annals evolved from that point till now? 
 Any suggestions you have for the journal, going forwards?
\item[Jon:]  The Annals has grown and changed quite a bit since I was co-editor with John Marden.   
John and I were receiving
about 300 papers per year and had initiated the possibility 
of electronic submission (which quickly became the norm).  
We had about 25 Associate Editors and were still using a data base system created by 
John Rice when he was a co-editor with Bernard Silverman.  
My understanding is that, now, the number 
of submissions has increased to around 700 per year, 
and the number of Associate Editors is up to about 50.   
The whole submissions and review process for the Annals and all the IMS journals 
is now handled through EJMS.  
Whereas the talk on the street in the early 2000's was of ``theory going away''
and the Annals of Statistics closing shop,  exactly the opposite has occurred!  
The era of ``big data'' and ``data science'' have created challenging new problems 
and created the need for many further theoretical developments to make sense of 
all the new methods being developed.  

The IMS has a history of creating new journals when the need arises:  the first IMS journal, 
the {\sl Annals of Mathematical Statistics}, was the sole 
IMS research journal until 1972 or 1973 when we created 
two new journals, the {\sl  Annals of Statistics} and the 
{\sl Annals of Probability}.   Jack Hall was chair
of the IMS Committee which recommended this split;    
Ingram Olkin was the inaugural editor of the {\sl Annals of Statistics}, and 
Ron Pyke was the inaugural editor of the {\sl Annals of Probability}.    
The next split came in (1991) when the IMS created the 
{\sl Annals of Applied Probability} (with J. Michael Steele as inaugural editor).
A further split came in (2007) when the IMS created the {\sl Annals of Applied Statistics} 
(with Bradley Efron as inaugural Editor-In-Chief, 
and three different area editors).
Perhaps the time has come for yet another new journal in the area of Data Science.  
In fact an IMS Committee (with Liza Levina as chair) is studying this possibility now.

\bigskip

\item[Mouli:] 
You have also served as IMS President very recently. How was that experience for you? 

\item[Jon:] It has been enlightening, rewarding, and considerably more work than I had anticipated.  
Since I am still somewhat ``in the harness'' as Past President  (at least until the Annual Meeting
this year in Vilnius), I won't say any more about that right now.  

\bigskip
\item[Mouli:] The field of statistics has clearly changed a lot since you started your career. In particular, it 
 now falls under the bigger umbrella of data science along with certain streams of engineering 
 and applied mathematics. In one sense, this is good 
as it enhances synergies and scope. On the other, there is also the possibility of a loss of 
identity. Any thoughts on what an 
optimal course for the discipline would look like? More generally, what are your 
perceptions of the discipline as you see it, today? 

\item[Jon:]  As a discipline or field, statistics is still fairly young, and it has indeed 
changed quite a lot during the span of my career.  The primary driver of this has been the 
enormous changes in computing power which have occurred over that time span.   
Statistics clearly needs to keep working to not only provide new methods for the many new
applications arising in various fields of science, but also ways of understanding the 
properties of the new methods.  This is likely to require quite a lot of new mathematics as 
well as new statistics and new ways of organizing statistical theory to tackle the new problems.  
As a discipline we need to be open to different ways that individuals and groups can contribute
to research.  
\bigskip
\item[Richard:]
One topic that we have had conversations on a number of times is `reproducible research', 
which is quite critical to keep the discipline on a solid honest footing. 
Do you think statisticians are meeting the bar when it comes to this in general? 

\item[Jon:] No, probably not yet.   ``Reproducible'' has a number of possible meanings in the context 
of the field of statistics:  all the way from documenting programs so that individuals 
can replicate their own computations a few years after completing them, to the conduct of scientific investigations
in a way that leads to the same conclusions from different labs or groups.  

\bigskip 
\item[Mouli:]  On July 30, 2010, on the occasion of your 65th birthday celebrations, 
you were made a `Knight of the Order of the 
       Netherlands Lion'. 
       Given your long term connections to and involvement with the Dutch School, this must have been 
special to you.  Perhaps you could tell us something about this?

\item[Jon:] Yes, that has been very special:  quite an unexpected honor.   
Willem van Zwet was the primary originator of this 
of course, but it certainly entailed support from many of my Dutch friends and 
collaborators,  Piet, Aad, Geurt, Chris, Richard, and more.  
As far as I know, 
Willem has organized Knighthoods for two statisticians in the US (Peter Bickel and myself), 
and two in Europe 
(Marie Huskova and Sara van de Geer).   
I confess I was caught completely off-guard when this happened in 2010, 
 but in retrospect I should have realized from Willem's hints that I should be 
 wearing a tie for the conference dinner  that something was up.   
 \bigskip
 
\item[Mouli:]  
Any thoughts of retirement? You appear to be enjoying your research as much as at any 
other time that we have known you! 

\item[Jon:]  
I am still enjoying research work quite a lot, and my intention is to continue 
that involvement for some time.   But my current plan is to retire from my teaching 
position at the UW in 2020.  I will take one more sabbatical leave during the Winter 
and Spring of 2019, then teach one more year (2019-2020) before retiring sometime 
between June and September 2020.  
\par\noindent
\end{description}

\begin{figure}[ht!]
\centering
\includegraphics[width=110mm]{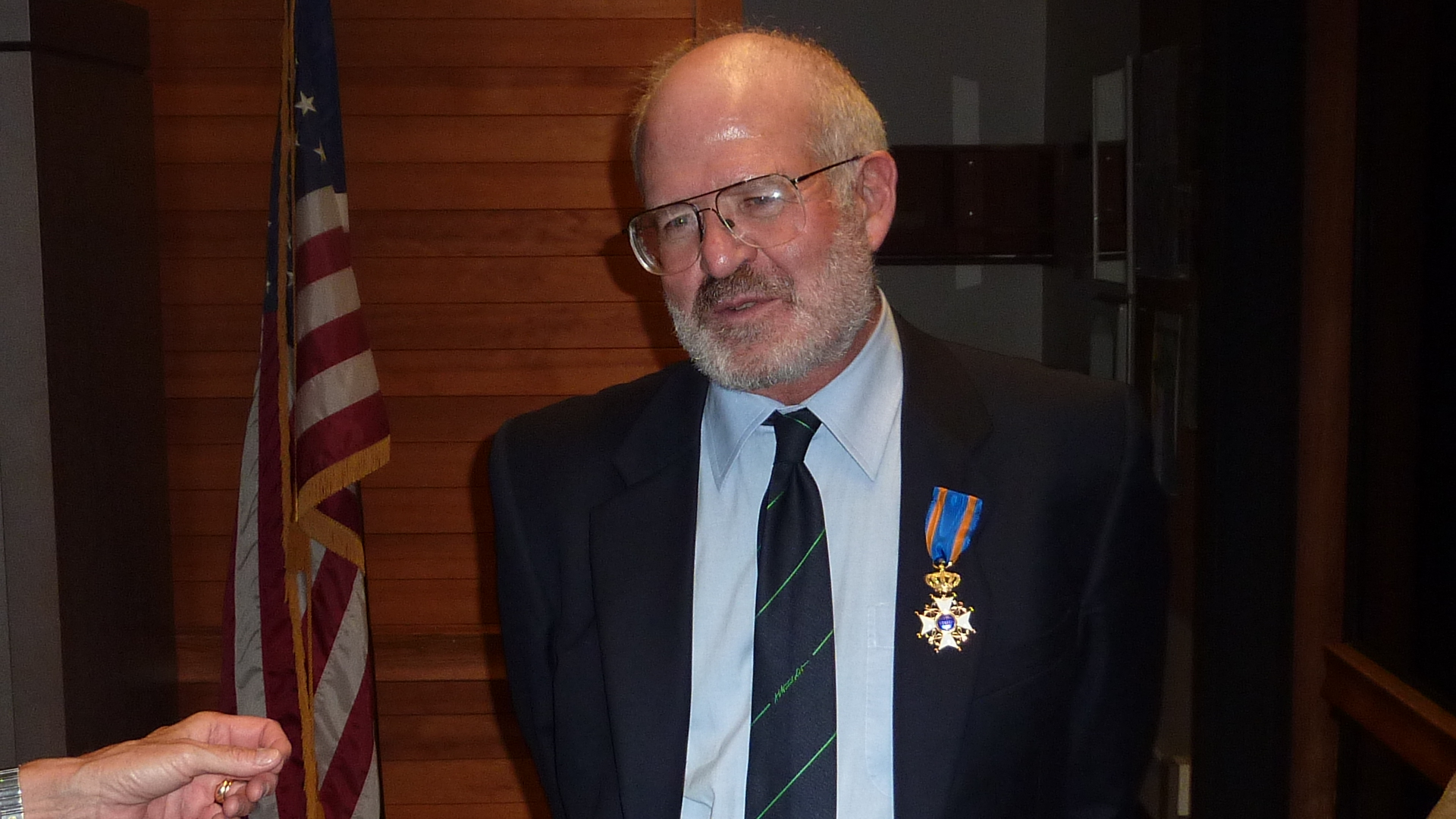}
\caption{Jon proudly wears the Emblem of the Order of the Netherlands Lion at UW, Seattle, 2010}
 \label{fig:figure7}
\end{figure}

\section{Personal Life and Interests}  
\begin{description}
\item[Mouli:] You were part of the Mountain Rescue Team in Seattle, something 
that dove-tailed nicely with your passion for mountaineering. 
I still remember the pager you used to carry around in the department. 
Tell us something about your experiences as part of that team. 

\item[Jon:]   I got involved in Seattle Mountain Rescue through Vic Ericson, 
the brother of a friend from my time in the Army and Vietnam, Paul Ericson.
Vic and I climbed together quite a bit during my graduate school days, and 
then again at the Gunks in New York during the time I was in Rochester 
and he was with ATT in  New Jersey. Vic started  working as a lobbyist for PNW Bell 
when he returned to Seattle, and he got roped into SMR by another lobbyist 
and long-time SMR member, Bill Robinson.

The missions with SMR varied enormously,
from searches for missing children, to rescues of people with broken legs, and 
to straightforward body recovery situations.   It frequently involved a push to get 
to victims of an accident as rapidly as possible.   I found it to be a challenging 
activity in which one could sometimes make a real impact in terms of getting an injured
person or party to safety.  The most satisfying missions involved actually getting someone
who had been injured or lost out safely.  
Serving as a rescue member of SMR involved quite a bit training and practice time (learning 
the rigging systems and relearning first aid and communication skills), but  with 
a committed group of people who were often quite different from my 
academic colleagues.  I participated actively as a ``rescue member'' of the group
from the mid-1980's until the late 1990's, and was involved in about 50 missions over
that time period.  
I also edited the newsletter, the ``Bergtrage'',  for SMR for  about 10 years.
\bigskip
\item[Richard:] 
You met your wife Vera through Mountain Rescue, isn't that correct? 
And you both share an avid passion for mountaineering! 

\item[Jon:] \ \  Yes;  Vera joined SMR just a little before I did in the mid 1980's.  Her connection
was through the Climbing Committee for the Seattle Mountaineers.  (The Climbing Committee 
is the group within the Mountaineers that organizes the Mountaineers'	 
climbing courses.)   We met 
on a mission to search for a missing skier at the White Pass ski area during 
March, 1986.   During the drive down to White Pass we had time to discuss the 
pros and cons of the types of climbing we each enjoyed the most:  she was into 
climbing elegant lines on solid rock, and had made several trips to Yosemite with friends
from the Mountaineers, while I was focussed more on peak bagging (which can involve
inelegant lines with lots of unstable loose rock).

\bigskip
\item[Richard:] 
You have been to Nepal several times. 
And some of these trips were with Norm Breslow. Would you recount some of your 
experiences in Nepal?

\begin{figure}[ht!]
\centering
\includegraphics[width=110mm]{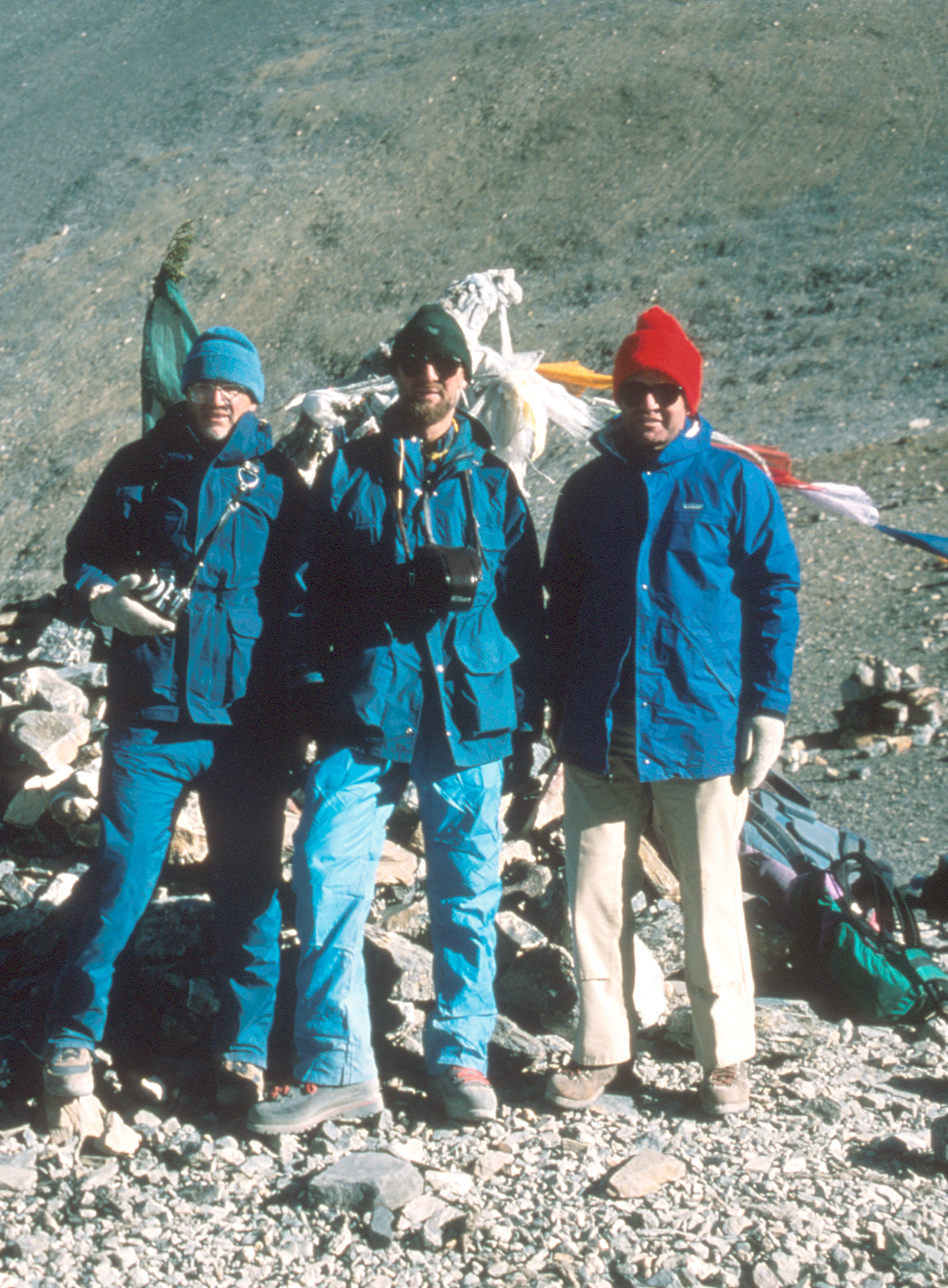}
\caption{David Thomas, Norm Breslow, and Jon at Thorong La (1989)}
\label{fig:figure8}
\end{figure}

\begin{figure}[ht!]
\centering
\includegraphics[width=110mm]{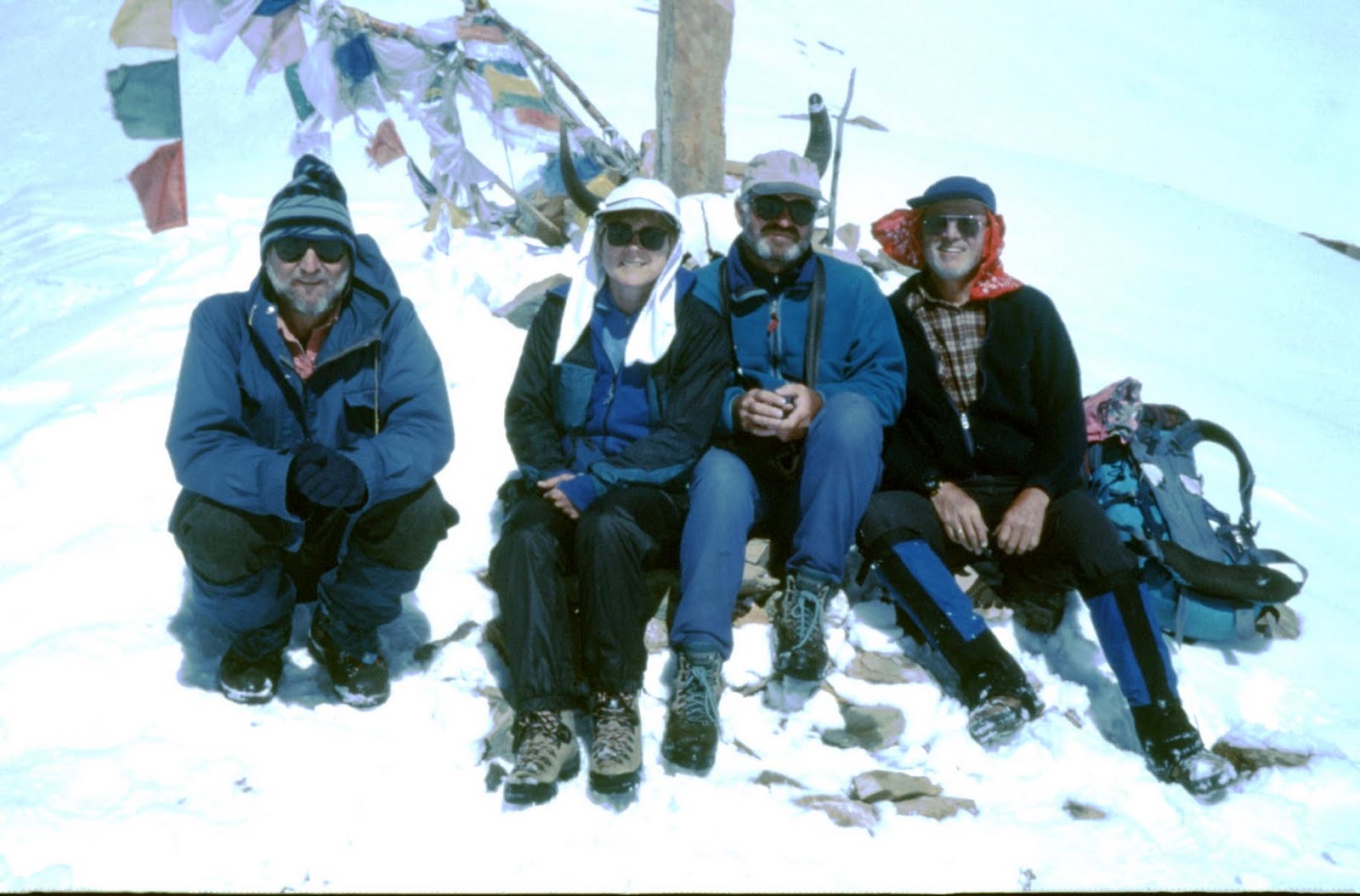}
\caption{Norm, Vera, Jon, and David up in the Himalayas at the Kagmara La (1999)}
\label{fig:figure9}
\end{figure}

\item[Jon:]  All of the Nepal trips were with Norm.
 The trips to Nepal in 1989, 1996, and 1999 were wonderful experiences.    
The first trip (1989) was with Norm Breslow and David Thomas, and 
involved a ``tea-house'' trek around Annapurna
over a period of three weeks, with one high pass, the Thorong La, at an elevation of 17,669 feet.   
On that trip, Norm visited the site of an enormous avalanche on the 
shoulder of Dhaulagiri which killed a Stanford classmate, Bill Ross, in 1969.  
(See ``American Dhaulagiri Expedition - 1969''. 
American Alpine Journal. American Alpine Club. 17 (1): 19. 1970.) 
David Thomas was involved in eradicating smallpox early in his career as an epidemiologist, a project
which took him to India, Pakistan and other countries in Asia and the Middle East.   

The second trip in 1996 was with Norm Breslow, another ``Reedie'' Peter Renz (both Norm and Peter did their 
undergraduate studies at Reed College in Portland, Oregon),  
and Rob Schaller, a friend of Peter's who
had been involved in an effort to set up a 
nuclear-powered surveillance device on top of Nanda Devi,  near the Indian border with China, in 1965. 
The goal on the 1996 trip was to climb a popular ``trekking peak'',  Mera Peak (6476m / 21246ft) 
to the south of Mount Everest in the Khumbu region of Nepal.  
To acclimate gradually we started the trek in Phaplu;  
the first few days were extremely wet with repeated close encounters with leeches and the wettest 
tenting conditions I have ever experienced.  As we got over the pass to the east of the 
Duhd-khosi river things started to dry out.
We ended up not quite making it all the way to the top of Mera, but thanks to an extremely 
fit pair of young porters who stomped 
uphill through new snow for hours, we did get quite high.   

The third trip, in 1999, was with Norm, David Thomas, and Vera, and took us on a three week trip 
into far less traveled country in Dolpo in western Nepal.  
In the course of those three weeks we 
went over three high passes and spent time near Lake Phoksundo 
where Vera fell in love with a young girl (Sangmu Royaka) whom 
we ended up supporting through school in Dunai and later in Kathmandu.  
Quite a magical trip all in all. 

\item[Mouli:] I don't know if you remember, but you and Vera and I were on the same flight from Seattle 
to Tokyo on that 1999 trip of yours, by coincidence! I was going to India. 
\bigskip
\item[Mouli:]  
Have you hiked in the Alps? Any other mountain ranges?

\item[Jon:] 
Yes, a bit.  But most of my climbing has been in North America: 
the Cascades and Tetons in the US,  and the British Columbia Coast 
Mountains in the vicinity of Mt. Waddington in Canada.  
\bigskip
\item[Richard:]
Do you still pursue mountain-climbing a bit?   And skiing?  

\item[Jon:]  \ \ Two hip replacement surgeries have slowed me down 
on this front a bit, but I am still doing some skiing and I hope to do more 
climbing when I retire.  Perhaps I will start working on climbing the peaks 
in the ``Bulger List'' that I haven't yet climbed.\\
See http://www.peakbagger.com/list.aspx?lid=21303.
 
\bigskip
\item[Mouli:] 
What other interests and hobbies do you have?

\item[Jon:] \ \  
Vera is getting me back into photography.  
The new digital cameras have phenomenal capabilities, and perhaps I can 
still learn how to program one of these gadgets!
\bigskip

\end{description}


\begin{description}
\item[Richard:] 
Thanks, Jon, for that fascinating insight into your life and career. 
It's clear you've led an active life, both professionally and personally, 
with more to come on both fronts in the future.

\item[Mouli:] 
Couldn't agree more with Richard! 
Thanks for letting us interview you, Jon, and all our very best for the coming years! 

\end{description}

\par\noindent 
{\bf References:}
\begin{itemize}
\item
Chernozhukov, V., Galichon, A., Hallin, M., and Henry, M. (2017).
Monge-Kantorovich depth, quantiles, ranks, and signs. 
{\sl Ann. Statist.} {\bf 45}, 223 - 256.
\item
Koltchinskii, Vladimir; 
Nickl, Richard; 
van de Geer, Sara; 
Wellner, Jon A. (2016).
The mathematical work of Evarist Gin\'e. 
{\sl Stochastic Process. Appl.} {\bf 126}, 3607 - 3622.  
\item
Chernoff, H.  (1964). 
 Estimation of the mode.  
 {\sl Ann. Inst. Statist. Math.} {\bf  16},  31 - 41.
\item
Gill, R. D.  (1989).
``Non- and semiparametric maximum likelihood estimators and the von Mises method (Part I)''.  
{\sl Scand. J. Statist.} {\bf 16}, 97 - 128.
\item
American Dhaulagiri Expedition - 1969. 
{\sl American Alpine Journal},  American Alpine Club. {\bf 17 (1)}, 19. 1970.
\item
Gardner, R. J. (2002).
The Brunn-Minkowski inequality. 
{\sl Bull. Amer. Math. Soc.} {\bf 39}, 355 - 405.
\item
Groeneboom, P. (1987).  Asymptotics for interval censored observations.
Report 87-18.  Department of Mathematics, University of Amsterdam.
\item
Groeneboom, P.  (1989).  
Brownian motion with a parabolic drift and Airy functions, 
{\sl Probab. Theory Related Fields} {\bf  81},  79 - 109.
\item
Groeneboom, P., Lalley, S.,  Temme, N. (2015).  
Chernoff's distribution and differential equations of parabolic and Airy type.
{\sl J. Math. Anal. Appl.} {\bf 423}, 1804 - 1824. 
\end{itemize}

\end{document}